\documentclass[11pt]{article}
\newsavebox{\PSLASH}
\sbox{\PSLASH}{$p$\hspace{-1.8mm}/}

\begin{document}
\title{\large \bf Matching LTB and FRW spacetimes through a null hypersurface}
\author{ S. Khakshournia$^{1}$
\footnote{Email Address: skhakshour@aeoi.org.ir} and R.
Mansouri$^{2,3}$ \footnote{Email Address: mansouri@ipm.ir}\\
$^{1}$Nuclear Science and Technology Research Institute (NSTRI),\\
Atomic Energy Organization of Iran, Tehran, Iran\\
$^{2}$Department of Physics, Sharif University of
Technology, Tehran, Iran\\
$^{3}$Institute for Studies in Physics and Mathematics (IPM), Tehran, Iran\\
}\maketitle
\[\]

\begin{abstract}

Matching of a LTB metric representing dust matter to a background
FRW universe across a null hypersurface is studied. In general, an
unrestricted matching is possible only if the background FRW is
flat or open. There is in general no gravitational impulsive wave
present on the null hypersurface which is shear-free and
expanding. Special cases of the vanishing pressure or energy
density on the hypersurface is discussed. In the case of vanishing
energy momentum tensor of the null hypersurface, i.e. in the case
of a null boundary, it turns out that all possible definitions of
the Hubble parameter on the null hypersurface, being those of LTB
or that of FRW, are equivalent, and that a flat FRW can only be
joined smoothly to a flat LTB.
\end{abstract}
\newpage

\section{Introduction}

All our observation in cosmology is along our past null-cone. What
if we implement the local inhomogeneities we observe in a globally
homogeneous universe in accordance with the cosmological principle
\cite{mansouri05}? Assume now a model universe in which the
background outside our past light cone is smoothed out to a
Fridmann-Robertson-Walker (FRW) universe, but on and within the
light cone in our vicinity the spacetime is inhomogeneous
\cite{khosravi07}. As a simple case we assume the inhomogeneous
part be given by a Lemaitre-Tolman-Bondi (LTB) solution to the
Einstein equations. Hence, the separating surface of the two
spacetime will become a null surface for the LTB observers. Now,
the question is if such a matching is exactly doable within the
general relativity assuming Einstein dynamics? We therefore
formulate the general problem of gluing a LTB metric to a FRW one
along a null hypersurface. Similar problem for the case of a
timelike hypersurface has already been studied
\cite{khakshournia02}. The null hypersurface is however more
delicate as we will see in this paper, at the same time it
has more cosmological applications.\\
For this purpose we use Barrab\`{e}s-Israel (BI) null shell
formalism \cite{Bar1} to investigate the matching and find the
junction conditions. Section 2 is devoted to the formulation of
the problem. The junction conditions are then explicitly written
in section 3. Section 4 is devoted to the discussion of the
impulsive gravitational wave on the null hypersurface and the
Penrose classification of the hypersurface for different cases.
Different applications and special cases are then discussed in
section 5. A conclusion follows then in section 6.\\
\textit{Conventions.}  Natural geometrized units, in which $G=c=1$
are used throughout the paper. The null hypersurface is denoted by
$\Sigma$. The symbol $|_{\Sigma}$ means "evaluated on the null
hypersurface ". We use square brackets [F] to denote the jump of
any quantity F across $\Sigma$. Latin indices range over the
intrinsic coordinates of $\Sigma$ denoted by $\xi^{a}$, and Greek
indices over the coordinates of the 4-manifolds.

\section{Formulation of the gluing LTB to FRW}

Consider an inhomogeneous metric containing dust matter
represented by a LTB cosmological model glued to a FRW background
universe along a null hypersurface. We choose the LTB metric to be
written in the synchronous comoving coordinates in the form
\cite{Bondi}
\begin{equation}\label{metricLTB}
ds^{2}_{-}=-dt_{-}^{2}+\frac{{R'}
^{2}}{1+E(r_{-})}dr_{-}^{2}+R^{2}(t_{-},r_{-})(d\theta^{2}
+\sin^{2} \theta d\varphi^{2}),
\end{equation}
where the overdot and prime denote partial differentiation with
respect to $t_{-}$ and $r_{-}$, respectively, and $E(r_{-})$ is an
arbitrary real function such that $E(r_{-})>-1$, and we take
$R'>0$ and $R > 0$ to avoid shell crossing and shell focusing,
respectively, of the dust matter during their radial motion. Then
the corresponding Einstein field equations turn out to be
\begin{eqnarray}\label{field}
\dot{R}^{2}(t_{-},r_{-})&=&E(r_{-})+\frac{2M(r_{-})}{R} ,\\
\hspace*{0.6cm}4\pi
\rho_{L}(t_{-},r_{-})&=&\frac{M'(r_{-})}{R^{2}R'},
\end{eqnarray}
where $\rho_{L}$ is the energy density and $M$ is another
arbitrary function. We take a FRW background universe described by
the following metric
\begin{equation}\label{metricFRW}
ds^{2}_{+}=-dt_{+}^{2}+\frac{a^2(t_{+})}{1-kr^{2}_{+}}dr_{+}^{2}+
r_{+}^2a^{2}(t_{+})(d\theta^{2}+\sin^{2} \theta d\varphi^{2}),
\end{equation}
where $k=+1, 0, -1$ distinguishes the closed, spatially flat, and
the open cosmological models, respectively. To glue the LTB
inhomogeneous patch and FRW background universe along the null
hypersurface $\Sigma$ we need to have
\begin{equation}\label{trans1}
r_{-}=r_{-}(t_{-}),\hspace{1cm}\frac{dr_{-}}{dt_{-}}=
\frac{\sqrt{1+E}}{R'(t_{-},r_{-})}\big|_{\Sigma},
\end{equation}
in the minus coordinates, and
\begin{equation}\label{trans2}
r_{+}=r_{+}(t_{+}),\hspace{1cm}\frac{dr_{+}}{dt_{+}}=
\frac{\sqrt{1-kr^{2}_{+}}}{a(t_{+})}\big|_{\Sigma},
\end{equation}
in the plus coordinates. Now, the requirement of the continuity of
the induced metric on $\Sigma$ yields the following matching
conditions:
\begin{eqnarray}\label{intrinmatch}
t_{+}=t_{-}=t,\hspace*{0.4cm}r_{+}a\stackrel{\Sigma}{=}R,\hspace*{0.4cm}
\frac{dr_{+}}{dr_{-}}&=&\frac{R'\sqrt{1-kr^{2}_{+}}}{a\sqrt{1+E}}\big|_{\Sigma}.
\end{eqnarray}
For further applications, we note that the differentiation of
$r_{+}a = R$ on $\Sigma$ leads to
\begin{equation}\label{junctioncond}
\sqrt{1-kr^{2}_{+}}+RH\stackrel{\Sigma}{=}\sqrt{1+E}+\dot{R},
\end{equation}
where $H=\frac{1}{a}\frac{da}{dt_{+}}$ is the Hubble parameter of
the FRW background. Taking $\xi ^{a}=(t,\theta,\varphi)$ with
$a=1,2,3$ as the intrinsic coordinates on $\Sigma$ we must
calculate the tangent basis vectors $e_{a}=\partial/\partial \xi
^{a}$ on both sides of $\Sigma$. Having written $x^{\mu}_{+}$ in
terms of $x^{\mu}_{-}$ by Eq. (\ref{intrinmatch}), we get
\begin{eqnarray}\label{verbien}
e^{\mu}_{t}|_{-}&=&\left(1,\frac{\sqrt{1+E}}{R'},0,0\right)
\big|_{\Sigma},\hspace*{1.4cm}
e^{\mu}_{\theta}|_{-}=\delta^{\mu}_{\theta},
\hspace*{1.6cm}e^{\mu}_{\varphi}|_{-}=\delta^{\mu}_{\varphi},\\
e^{\mu}_{t}|_{+}&=&\left(1,\frac{\sqrt{1-kr^{2}_{+}}}{a},0,0\right)\big|_{\Sigma},
\hspace*{0.6cm}e^{\mu}_{\theta}|_{+}=\delta^{\mu}_{\theta},
\hspace*{0.6cm}e^{\mu}_{\varphi}|_{+}=\delta^{\mu}_{\varphi}.
\end{eqnarray}
Let $t$ be a parameter on the null geodesic generators of
$\Sigma$, we choose the tangent-normal vector $n^{\mu}$ to
coincide with the tangent basis vector associated with the
parameter $t$, so that $n^{\mu} = e^{\mu}_{t}$. We may then
complete the basis by a transverse null vector $N^{\mu}$ uniquely
defined by the four conditions $n_{\mu}N^{\mu} = -1$,
$N_{\mu}e^{\mu}_{A} = 0$ $(A = \theta,\varphi)$, and
$N_{\mu}N^{\mu} = 0$. We find
\begin{eqnarray}\label{normaltrans}
N_{\mu}|_{-}&=&\frac{1}{2}\left(-1,\frac{-R'}{\sqrt{1+E}},0,0\right)\big|_{\Sigma},\\
N_{\mu}|_{+}&=&\frac{1}{2}\left(-1,\frac{-a}{\sqrt{1-kr^{2}_{+}}},0,0\right)\big|_{\Sigma}.
\end{eqnarray}
Furthermore, the induced metric on $\Sigma$ given by $g_{ab} =
g_{\mu\nu}e^{\mu}_{a}e^{\nu}_{b}|_{\pm}$ is computed to be\\
$g_{ab}$ = diag$\left(0,R^{2},R^{2}\sin^{2}\theta\right)$, which
is the same on both sides of the hypersurface. Defining a
pseudo-inverse of the induced metric $g_{ab}$ on $\Sigma$ as
$g_{*}^{ac}g_{bc} = \delta_{b}^{a}+ n^{a}N_{\mu}e^{\mu}_{b}$, with
$n^{a} = \delta^{a}_{t}$ \cite{Bar1}, one
gets $g^{ab}_{*}$ = diag$\left(0,\frac{1}{R^{2}},\frac{1}{R^{2}\sin^{2}\theta}\right)$.\\
The final junction condition is formulated in terms of the jump in
the extrinsic curvature. Using the definition ${\cal K}_{ab} =
e^{\mu}_{a}e^{\nu}_{b}\nabla_{\mu}N_{\nu}$, we may therefore
compute the transverse extrinsic curvature tensor \cite{Bar1} on
both sides of $\Sigma$. Its non-vanishing components on the minus
side are found as
\begin{equation}\label{Ktetteta1}
{\cal K}_{\theta\theta}|_{-}=\sin^{-2}\theta{\cal
K}_{\varphi\varphi}|_{-}=\frac{R}{2}(\dot{R}-\sqrt{1+E})\big|_{\Sigma},
\end{equation}
\begin{equation}\label{Krr1}
{\cal K}_{tt}|_{-}=H_{R'}\big|_{\Sigma},
\end{equation}
where we have defined $H_{R'} = \frac{\dot R'}{R'}$  as one of the
possible definitions of the Hubble function for a LTB metric while
the other possibility is $H_R = \frac{\dot R}{R}$. Although these
definitions coincide in the case of FRW, they are different for
LTB. The corresponding non-vanishing components on the plus side
are
\begin{equation}\label{Ktetteta2}
{\cal K}_{\theta\theta}|_{+}=\sin^{-2}\theta{\cal
K}_{\varphi\varphi}|_{-}=\frac{R}{2}(RH-\sqrt{1-kr^{2}_{+}})\big|_{\Sigma},
\end{equation}
\begin{equation}\label{Krr2}
{\cal K}_{tt}|_{+}=H.
\end{equation}
The jump in the transverse extrinsic curvature across the null
hypersurface, given by $\gamma_{ab}=2[{\cal K}_{ab}]$, has the
following non-zero components:
\begin{equation}\label{gamma22}
\gamma_{\theta\theta}=\sin^{-2}\theta\gamma_{\varphi\varphi}=
2R(\sqrt{1+E}-\sqrt{1-kr^{2}_{+}})\big|_{\Sigma},
\end{equation}
\begin{equation}\label{gamma11}
\gamma_{tt}=2\left(H-H_{R'}\right)\big|_{\Sigma},
\end{equation}
where we have used Eq. (\ref{junctioncond}).

\section{Junction conditions}

The surface energy-momentum tensor of the lightlike shell having
the null hypersurface $\Sigma$ as its history is directly related
to the jump in the transverse extrinsic curvature.  It can be
expressed in the intrinsic coordinates $\xi^{a}$ as follows
\cite{Pois}
\begin{equation}\label{juncnullintr}
S^{ab}=f n^{a}n^{b}+pg_{*}^{ab}+j^{a}n^{b}+j^{b}n^{a},
\end{equation}
where
\begin{equation}\label{nullenergy}
f=-\frac{1}{16\pi}g_{*}^{ab}\gamma_{ab}
\end{equation}
represents the surface energy density,
\begin{equation}\label{nullpressure}
p=-\frac{1}{16\pi}\gamma_{ab}n^{a}n^{b}
\end{equation}
displays the isotropic surface pressure, and
\begin{equation}\label{nullcurrent}
j^{a}=-\frac{1}{16\pi}g_{*}^{ac}\gamma_{cd}n^{d}
\end{equation}
represents the surface current of the lightlike shell. All these
surface quantities are measured by a family of freely-moving
observers crossing the null hypersurface. Using the jumps in the
extrinsic curvature obtained above, we notice first that the
surface current term given by (\ref{nullcurrent}) vanishes
identically. From Eqs. (\ref{nullenergy}) and (\ref{nullpressure})
the energy density and pressure are then calculated as
\begin{eqnarray}\label{densnull}
16\pi f = -\frac{2}{R^{2}}\big|_{\Sigma}\gamma_{\theta\theta}=
\frac{4}{R}(\sqrt{1-kr^{2}_{+}}-\sqrt{1+E})\big|_{\Sigma},
\end{eqnarray}
\begin{eqnarray}
\label{pressrnull} 16\pi p = -\gamma_{tt}=
2\left(H_{R'}-H\right)\big|_{\Sigma}.
\end{eqnarray}
Assuming the positivity of the surface energy density of the shell
given by (\ref{densnull}) we see that the matching of an LTB
metric along a null hypersurface to the FRW metric is possible
only if
\begin{equation}\label{constraint}
E(r_{-})\stackrel{\Sigma}{\leq} -kr^{2}_{+}.
\end{equation}
Taking into account that there is always $-1 < E$, the following
cases are possible:
\begin{eqnarray}\label{k-cases}
k& =& 0  \hspace*{1.2 cm}   \Rightarrow \hspace*{1cm}
E = 0\hspace*{1 cm} and \hspace*{1 cm}-1 < E < 0, \\
k &=& + 1 \hspace*{1 cm} \Rightarrow \hspace*{1 cm} -1 < E \leq -r^{2}_{+},  \\
k &=& -1  \hspace*{1 cm} \Rightarrow \hspace*{1 cm} E = 0,
\hspace*{1 cm} -1 < E < 0  \hspace*{1 cm}and \hspace*{1 cm}0 < E
\leq r^{2}_{+}.
\end{eqnarray}
Note that all these constraints are valid on the light cone, i.e.
both $r_-$ and $r_+$ may, in general, take different values
between zero and infinity, except for the FRW case with $k = +1$,
where we must have $0 < r_{+} < 1$. Therefore, cases bounded by
$-1< E \leq -r^{2}_{+}$ and $0 < E < r^{2}_{+}$ may not be valid
globally. We will call such cases restricted matching. Therefore,
to have an unrestricted matching, the background FRW has to be
flat or open. In each case the LTB may then be flat or closed. It
should be noted that for the case of a marginally bound (flat)
inhomogeneous patch $(E=0)$ glued to a flat FRW background
$(k=0)$, using Eq. (\ref{densnull}), we find that $f=0$ so that
the null hypersurface is the history of a lightlike shell with the
only possibility of admitting a surface pressure. From
(\ref{pressrnull}) we also see that the LTB Hubble function
$H_{R'}$ may be bigger or smaller than the FRW Hubble parameter
$H$ on the hypersurface $\Sigma$, depending on the surface
pressure on $\Sigma $ being negative or
positive.\\

\section{The case of gravitational impulsive wave}

Gluing of manifolds along null hypersurfaces is a tricky action.
Although similar expression to (\ref{juncnullintr}) in the case of
time-like or space-like hypersurfaces includes all the information
about the junction, this is not so for the null case. In general,
there is a part of $\gamma_{ab}$, denoted by $\hat{\gamma}_{ab}$,
which does not contribute to the expression (\ref{juncnullintr})
for the intrinsic energy-momentum tensor $S^{ab}$ of the null
shell, and is interpreted as an impulsive gravitational wave
propagating independently of the null shell. The expression for
$\hat{\gamma}_{ab}$ is given by \cite{Bar4}
\begin{equation}\label{gammachek}
\hat{\gamma}_{ab}=\gamma_{ab}-\frac{1}{2}g_{*}^{cd}\gamma_{cd}g_{ab}+2
n^{c}\gamma_{c(a}N_{b)}+\gamma_{cd}n^{c}n^{d}N_{a}N_{b},
\end{equation}
where $N_{a}=N_{\mu}e^{\mu}_{a}=(-1,0,0)$. In our case, however,
it can easily be shown that all components of $\hat{\gamma}_{ab}$
defined vanish identically. This is a hint that there is no
gravitational impulsive wave across the null
hypersurface of the junction between LTB and FRW. \\

The absence of gravitational waves having the null hypersurface
$\Sigma$ as history can explicitly be seen in the following way.
Let us first construct a null tetrad frame on $\Sigma$. Consider a
congruence of timelike geodesics with continuous 4-velocity $u$
across $\Sigma$ so that
$[u_{\alpha}u^{\alpha}]=[u_{\alpha}e^{\alpha}_{a}]=0$. On the null
hypersurface $\Sigma$ we have the normal $n^{\mu}_{-}$, which is
tangential to $\Sigma$, and the timelike vector field $u^{\mu}$
tangent to the matter world lines in the LTB manifold and FRW
background crossing the null hypersurface such that
$u^{\alpha}n_{\alpha}=-s<0$. It is then advantageous to introduce
on $\Sigma$ a transverse null vector field $l^{\mu}$ defined by
$l^{\mu}=-\frac{1}{2s^{2}}n^{\mu}+\frac{1}{s}u^{\mu}$, satisfying
the normalization condition $l^{\alpha}n_{\alpha}=-1$ \cite{Bar3}.
We find
\begin{equation}\label{l}
l^{\mu}=\frac{1}{2}\left(1,-\frac{\sqrt{1+E}}{R'},0,0\right)\big|_{\Sigma}.
\end{equation}
Let next $m^{\mu}$ be a complex covariant vector field and its
complex conjugate, $\bar{m}^{\mu}$, being chosen so that they are
null ($m^{\alpha}m_{\alpha}=\bar{m}^{\alpha}\bar{m}_{\alpha}=0$),
tangent to $\Sigma$, orthogonal to $n^{\mu}$ and $l^{\mu}$, and
satisfy $m_{\alpha}\bar{m}^{\alpha}=1$. Now $n^{\mu}_{-}$,
$l^{\mu}$, $m^{\mu}$, and $\bar{m}^{\mu}$ constitute the desired
null tetrad frame on $\Sigma$ which will be used in the following.
We get
\begin{equation}\label{m}
m^{\mu}=\frac{1}{R\sqrt{2}}
\left(0,0,i,\frac{1}{\sin\theta}\right)\big|_{\Sigma},
\end{equation}
\begin{equation}\label{mbar}
\bar{m}^{\mu}=\frac{1}{R\sqrt{2}}
\left(0,0,-i,\frac{1}{\sin\theta}\right)\big|_{\Sigma}.
\end{equation}
Using this null tetrad, the Newman-Penrose component of the
singular part of the Weyl tensor of Petrov type N characterizing
an impulsive gravitational wave with history $\Sigma$ is
calculated as \cite{Bar3}
\begin{eqnarray}\label{Psi}
\breve{\Psi}_{4}&=&\frac{1}{2}\gamma_{ab}\bar{m}^{a}\bar{m}^{b},\nonumber\\
&=&\frac{1}{2}(\gamma_{\theta\theta}(\bar{m}^{2})^{2}+
\gamma_{\varphi\varphi}(\bar{m}^{3})^{2}),\nonumber\\
&=&0,
\end{eqnarray}
where we have used Eqs. (\ref{gamma22}) and (\ref{mbar}). This
shows explicitly that  there is no impulsive gravitational wave
present and the null hypersurface $\Sigma$ is just the history of
a lightlike shell of matter being characterized in general by the
surface energy density and isotropic pressure given by Eqs.
(\ref{densnull}), and (\ref{pressrnull}), respectively. In this
case the induced geometry on $\Sigma$, inherited from the
embedding spacetimes, is of type I according to a classification
introduced by Penrose \cite{Bar3,Penros}.\\
The expansion $\theta$ and complex shear $\sigma$ of the geodesic
generators of the null hypersurface $\Sigma$ can now be defined by
the following relations using the null tetrad \cite{Bar3}:
\begin{equation}\label{theta}
\theta=m^{\mu}\bar{m}^{\nu}\nabla_{\nu}n_{\mu}=
\frac{1}{R}(RH+\sqrt{1-kr^{2}_{+}})\big|_{\Sigma},
\end{equation}
\begin{equation}\label{sigma}
\sigma=m^{\mu}m^{\nu}\nabla_{\nu}n_{\mu}=0.
\end{equation}
The LTB inhomogeneous manifold is, therefore, joined to the
background FRW universe through a shear-free, expanding null
hypersurface
$\Sigma$.\\
We have, therefore, shown that the matching of a LTB metric to an
arbitrary FRW metric is possible given the constraint junction
(\ref{constraint}) is satisfied. Although the hypersurface of
junction is in general supporting energy and pressure, there is no
impulsive gravitational wave with the history $\Sigma$, which
turns out
to be shear-free but expanding.\\

\section{Application to Special Cases}

We have already seen that not all cases of matching a LTB to FRW
along a null hypersurface is possible. Now, we would like to check
different interesting cosmological cases which are possible,
depending on the different forms of the energy momentum tensor on
the hypersurface.

\vspace{0.5 cm}

 i) \textbf{Pressureless hypersurface}

\vspace{0.5 cm}

Let us here consider the case that there is no surface isotropic
pressure. Then it follows from (\ref{pressrnull}) that there is no
jump in the Hubble function across $\Sigma$: $H = H_{R'}$. In this
case the energy density $f$ is the only non-vanishing surface
quantity due to the presence of a lightlike shell of matter with
the history $\Sigma$. Now, we have
$\gamma_{a}=\gamma_{ab}n^{b}=\gamma_{tt}=0$. Therefore, the
induced geometry on $\Sigma$ is of type III, according to the
Penrose classification of induced geometries on  $\Sigma$
\cite{Bar3,Penros}. We may now write down the junction equation
relating the properties of $\Sigma$ to those of outside medium
described by the energy-momentum tensor $T^{\pm}_{\mu\nu}$
\cite{Bar3}:
\begin{equation}\label{nulljunction}
8\pi[T_{\mu\nu}n^{\mu}n^{\nu}]\stackrel{\Sigma}{=}\theta\gamma^{\dag},
\end{equation}
where $\gamma^{\dag}=\gamma_{a}n^{a}=0$ since the induced geometry
on $\Sigma$ is of type III. Then by noting that LTB and FRW are
perfect fluid spacetimes we see that Eq. (\ref{nulljunction}) is
reduced to the form
\begin{equation}\label{densityjump}
\rho_{L}\stackrel{\Sigma}{=}\rho_{b}+p_{b},
\end{equation}
where $\rho_{L}$ and $\rho_{b}$ are the energy densities in LTB
and FRW spacetimes respectively, and  $p_{b}$ is the fluid
pressure in FRW universe. Particularly, in the case of the
pressure-free FRW universe we see that having a pressureless shell
on the null hypersurface of junction requires that the jump in
the energy density across $\Sigma$ vanish.\\

\vspace{0.5 cm}

ii) \textbf{Vanishing the surface energy-momentum tensor: boundary
layer}

\vspace{0.5 cm}

Looking now for a special case of great cosmological interest,
i.e. smooth matching of an inhomogeneous patch to a homogeneous
FRW background, let in addition to the surface pressure $p$, the
surface energy density $f$ vanish too. From (\ref{constraint})
this requires that $E(r_{-})\stackrel{\Sigma}{=}-kr^{2}_{+}$. Then
in this case there would be neither a surface energy-momentum
tensor on $\Sigma$ nor the impulsive gravitational wave. We
therefore conclude that the null hypersurface $\Sigma$ is just a
smooth boundary, like our past like cone in observational
cosmology. On the other hand, since $\gamma_{\theta\theta}=0$ and
using Eqs. (\ref{gamma22}) and (\ref{junctioncond}), we find that
\begin{equation}
H \stackrel{\Sigma}{=} H_R \stackrel{\Sigma}{=}H_{R'},
\end{equation}
which is an interesting result which has been verified numerically
\cite{khosravi07}. It says that for the case of a smooth matching
of a LTB metric to a FRW background, the LTB Hubble functions on
$\Sigma$ related to the metric functions $R$ or $R'$ are identical
to the corresponding FRW Hubble function. We may therefore define
a Hubble parameter without any ambiguity. These results are valid
for any open or flat model, and in the
restricted matchings, also for closed ones. \\

\vspace{0.5 cm}

 iii) \textbf{Junction of the flat-flat case along a null
boundary hypersurface}

\vspace{0.5 cm}

 A more closed look at the relation (\ref{densnull})  shows that a flat or
open FRW (LTB) can only be matched smoothly to a flat or open LTB
(FRW), respectively. Therefore, if one is going to model the
actual local inhomogeneities of the universe using a LTB part
joined to a flat FRW along the past light cone, then it must also
be flat, otherwise there will be a null matter shell with the
history $\Sigma$.\\
The relation between $r_{-}$ and $r_{+}$ is an interesting
question giving an insight in the LTB metric, and best studied in
the flat-flat case. From (\ref{intrinmatch}) we infer for the
flat-flat case with $E(r) = 0 = k$
\begin{equation}
\frac{dr_{+}}{dr_{-}} = \frac{R'}{a}\big|_{\Sigma}.
\end{equation}
Now, let us see if one may have $r_{+} = r_{-}$ on the null
hypersurface. For this to be the case one must have $R'(r_{-},t)
\stackrel{\Sigma}{=} a(t)$. We know, however, that for the LTB
solution in the flat case the metric function is given by
\begin{equation}
R(r_{-},t) = (\frac{4\pi}{3}\rho_c)^{1/3} r_{-} (t -
t_{n}(r_{-}))^{ 2/3},
\end{equation}
where $\rho_c$ is just a constant and the $t_{n}$ is the so-called
bang time \cite{mansouri05,khosravi07}. Using this, from the above
condition, we obtain $t_n^{'}\big|_{\Sigma} = 0$. Therefore, the
derivative of the bang time must be zero all along the null
hypersurface, i.e. for all $r_{-}$. We then conclude that the bang
time must be a constant. This is equivalent to reducing LTB to a
FRW metric. Hence, for a flat LTB to be glued to a flat FRW the
comoving coordinate $r$ can not be the same on different sides of
the hypersurface. This may also be stated in the following way:
although on the null hypersurface the relation $R(r_-,t)
\stackrel{\Sigma}{=} a(t)r_{+}$ is valid, we always have
$R'\big|_{\Sigma} \neq a(t)$, except for the case of LTB being
reduced to FRW. This has explicitly been shown for some special
cases in \cite{khosravi07}. Therefore, for the junction of LTB to
FRW, we always have $r_{-} \neq r_{+}$ on the null hypersurface.
Obviously, this fact being proved for the flat-flat case, is also
true in general.

\section{Conclusion}

We have examined the matching of a LTB inhomogeneous solution to a
background FRW universe across a null hypersurface. In general,
the null hypersurface is the history of a lightlike matter shell
having surface energy density and isotropic pressure but no
impulsive gravitational wave. We have shown that the assumption of
positivity of the surface energy density of the null shell imposes
the criterion (\ref{constraint}) for gluing a LTB metric to the
FRW universe, with the result that an unrestricted
matching is only possible if the background FRW is open or flat. \\
If the surface pressure on the null hypersurface vanishes, then
the jump of the energy density across the null hypersurface is
related to the fluid pressure of the background FRW universe. The
case of vanishing both surface energy density and pressure on the
null hypersurface leads to the interesting result that the FRW
Hubble parameter and both definitions of the LTB Hubble parameters
are equivalent. Therefore, in the case of a surface boundary
junction a Hubble parameter may be defined without ambiguity. It
has also been shown that a flat FRW can only be joined to a flat
LTB along a null boundary.


\begin{thebibliography}{99}
\bibitem{mansouri05}
R. Mansouri, [arXiv:astro-ph/0512605].
\bibitem{khosravi07}
Sh. Khosravi, E. Kourkchi, R. Mansouri, and Y. Akrami,
[arXiv:astro-ph/0702282].
\bibitem{khakshournia02}
Khakshournia and R. Mansouri, Phys. Rev. D \textbf{65}, 027302,
(2002).
\bibitem{Bar1}
C. Barrab\`{e}s and W. Israel, Phys, Rev. D \textbf{43}, 1129,
(1991).
\bibitem{Bondi} G. Lemaitre, Ann. soc. Sci. Bruxelles Ser.1, A53, 51
(1933); R.C. Tolman,Proc. \\
\hspace*{0.22cm}Nat1. Acad. Sci. U.S.A. 20,410 (1934); H. Bondi,
Mon. Not. R. Astron. Soc.\\
\hspace*{0.19cm}107, 343 (1947).
\bibitem{Pois}
 E. Poisson, [arXiv:gr-qc/0207101].
\bibitem{Bar4}
C. Barrab\`{e}s, G. F. Bressange, and P. A. Hogan, Phys. Rev. D
\textbf{55}, 3477 (1997).
\bibitem{Bar3}
C. Barrab\`{e}s, P. A. Hogan, Phys. Rev. D \textbf{58}, 044013
(1998).
\bibitem{Penros}
R. Penrose, General Relativity, Clarendom, Oxford, p. 101 (1972).


\end{thebibliography}
\end{document}